# Energy Performance Analysis of Distributed Renewables: Pacific Northwest Smart Grid Demonstration

Donald J. Hammerstrom, *Senior Member*, *IEEE*, Yashodhan P. Agalgaonkar, and Stephen T. Elbert, *Member*, *IEEE*

*Abstract*— The **Pacific Northwest Smart Grid Demonstration (PNWSGD) was an electricity grid modernization project conducted in the Northwest U.S. This paper presents the PNWSGD's analysis of renewable generation at the Renewable Energy Park located in the City of Ellensburg, WA, one of 11 utilities that participated in the PNWSGD. The Renewable Energy Park consisted of grid-integrated distributed wind and photovoltaic (PV) renewable generation. The site is pioneering an interesting model for community investment in renewable energy generation. The energy yield of the resources was analyzed in order to assess each resource's performance, and the city's energy savings were estimated using the Bonneville Power Administration's (BPA) rate structure. This analysis also addressed the impact of the renewable resources on demand charges that are incurred most months by the city. The most important observation was that unanticipated wind tower and blade failures caused the city to remove the wind turbines, fearing that flying or falling debris might harm persons in the park. The energy from the PV arrays was reliable and offset both the city's energy costs and demand charges. Many of the small wind turbines failed to produce much energy or failed entirely during the short time they were installed. The community energy park concept is an intriguing model for community investment in renewable resources, but the lessons in this paper should be considered.**

*Index Terms*— Cost benefit analysis, Distributed power generation, Solar energy, Wind energy

## I. Introduction

The steady growth of renewables across the U.S. persists in order to achieve comprehensive next decade targets. By 2020, the goal is to achieve 100 MW of renewables across the federally subsidized housing and allocating public land for 10 GW renewable projects. Also by 2025, the renewable energy goal for military installations is 3 GW. Overall wind and photovoltaic (PV) electricity generation capacity is expected to double by then [1]. Along with integrating renewables, policy makers have found it essential to sensitize societal outlook towards the necessity of renewable energy. Scientists, utility engineers and policy makers have found a multi-objective concept of renewable energy parks quite attractive. A Renewable Energy Park is a planned generation capacity consisting of sources such as wind and PV. Along with generation of clean energy, its objective is to serve as an educational setup for schools, universities and also to promote eco-tourism. Typically, these energy parks consist of distributed PV and distributed wind technologies.

Distributed wind is specifically an important technology for the U.S. economy [2]. Distributed wind is utilized across the US in various applications: residential or on-grid power, industrial applications, or through community projects [3]. Most of the distributed wind technology in the U.S is grid integrated. The financial incentive received by distributed wind from various sources in fiscal year 2014 is $20.4 million. Distributed wind cumulative connected capacity in the US had reached around 906 MW by the end of 2014. Particularly in Washington state, the U.S. Department of Agriculture's Rural Energy funded 7 distributed wind projects during 2003 to 2014 costing $661,284. U.S. small wind turbine manufacturers are global leaders and export significant number of small wind turbines internationally. U.S. wind turbine manufacturers exported approximately 11 MW of distributed wind technology in 2014. Distributed wind was exported to various countries such as UK, Italy etc. The capacity factor for a sample of distributed wind projects installed in 2013 and 2014 is 25% [5].

Though this export market presents an opportunity for U.S. domestic manufacturers, this technology faces several challenges. One of the major challenges is comparison with the distributed PV technology. Furthermore, the low cost of electricity from other sources is also a challenge. In order to overcome these challenges, distributed wind depends upon a policy support. The U.S. Department of Energy (DOE) has launched several programs which fund technology advancement and competence improvements. Some manufactures have taken advantage of these programs to develop distributed wind technology further. From a policy making point of view, distributed wind faces several challenges, such as zoning issues and financial/energy performance [6]. Previously references [7] and [8] used the projected amount of electricity generation from distributed wind and carried out assessment of its futuristic performance. However, in order to evaluate realistic performance and study various challenges discussed above, energy performance analysis through practical measurements of existing distributed wind projects becomes essential.

This work was supported in part by the U.S. Department of Energy under Pacific Northwest Smart Grid Demonstration.

Y. P. Agalgaonkar, D. J. Hammerstrom, and S. T. Elbert are employees of Battelle Memorial Institute at the Pacific Northwest National Laboratory, Richland, WA 99354.

THE PAPER WAS DRAFTED IN 2016 FOR WIDER DISSEMINATION. THE ORIGINAL RESULTS WERE BRIEFLY MENTIONED IN PACIFIC NORTHWEST SMART GRID DEMONSTRATION PROJECT TECHNOLOGY PERFORMANCE REPORT VOLUME 1 PUBLISHED ON JUNE 21, 2015.
HTTPS://WWW.SMARTGRID.GOV/DOCUMENT/PACIFIC_NORTHWEST_SMART_GRID_TECHNOLOGY_PERFORMANCE.HTML

The City of Ellensburg Renewable Energy Park was one such research experiment conducted under the auspices of the Pacific Northwest Smart Grid Demonstration (PNWSGD) project that was co-funded by the U.S. DOE 2009-15. The City of Ellensburg installed distributed wind and PV generation assets at a distribution scale. At distribution scale, the monetary value of generated renewable energy lies primarily in the avoided energy purchases. Here, performance is studied based on the assessment of reduction in Bonneville Power Administration (BPA) Tier 2 expense. This paper reports various technical and policy related observations regarding the Renewable Energy Park and distributed wind based on this energy performance analysis.

This paper is organized as follows. Section II discusses the City of Ellensburg Renewable Energy Park in detail. Section III discusses the BPA tiered rate structure. Detailed discussion about energy performance of all the distributed renewable is presented in Section IV. Section V summarizes the analysis findings from a technical and policy perspective.

## II. ELLENSBURG RENEWABLE ENERGY PARK

The City of Ellensburg, Washington, is a municipality that serves about 10,000 electric and 5,500 gas customers. Through their participation in the PNWSGD Project, the city added both PV and wind generation capacity. The City of Ellensburg augmented their pre-exiting thin-film PV capacity by 40 kW and added 43 kW of nameplate wind capacity. Customer participation was one of the key features of this project. The community Renewable Energy Park consolidated citizens' efforts to test and use more renewable resources. Residents could buy into the project without having to construct and operate the generators themselves. The residents could take advantage of economies of scale by building larger and more cost-effective generators than they might construct on their own properties. The customers could participate with a minimal $250 investment. Furthermore, the city believed that the centralized park generation resource could be more safely managed. The city and its residents could learn about and compare the renewable generator technologies that will inform their future energy decisions [9]. The site is at a multipurpose community park that is publicly accessible. The diagram in Figure 1 labels the four installation phases for the arrays of solar panels and indicates the approximate locations of the solar arrays and nine wind turbines. The weather metrology tower was installed just beyond the upper right corner of this diagram.

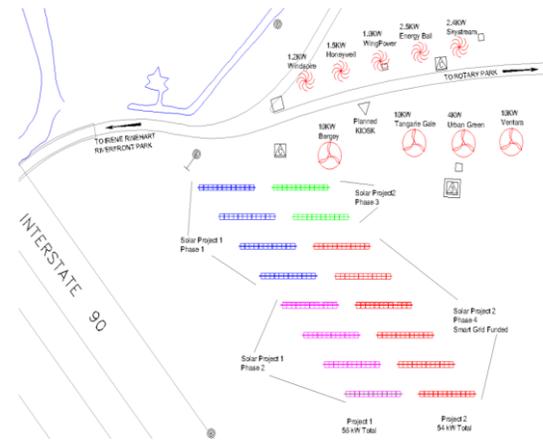

Figure 1. Layout of the Renewable Energy Park

The array on the right is the set of thin-film panels. From left to right, the four larger commercial-class wind turbines south of the path are Bergey, Tangarie, Urban Green Energy, and Ventera Wind, and the five smaller residential-class wind turbines north of the path are the Windspire, Honeywell WindTronics, Wing Power, Energy Ball, and Southwest Windpower Skystream wind turbines (Presently, Xzeres Wind).

The city had intended to make the data created by the project at the renewable park available to its residents, to researchers at Central Washington University, located in Ellensburg, and even to local teachers and their K–12 curricula. The City of Ellensburg received some qualitative value from its very visible investment in green energy resources and describes the park as an "eco-tourism" site [9].

The City of Ellensburg installed metrology equipment and made SCADA and other general site improvements through their participation in the PNWSGD. The major focus of their project participation was the installation and testing of PV generation systems, five residential-class wind turbine systems, and four larger commercial-class wind turbine systems. In order to carry out energy performance analysis for these renewables, BPA tiered rates need to be considered. The following section introduces BPA rate structure.

## III. BPA TIERED RATES

The monetary value of renewable distributed energy lies primarily in the displacement of electrical energy, avoided power energy purchases as well as mitigating system peak power capacity and avoiding consequent demand charges that are based, in part, on the maximum energy is that consumed by the city during an hour each month. The value of energy and peak capacity at utilities that are BPA customers may be evaluated based on BPA Tiered Rates Methodology [10]. The differential monetized values of energy and capacity at these utilities is estimated well from BPA load-shaping rates and demand rates. A time-of-use type of price signal is sent to BPA customers by the differentiation of heavy-load hours (HLHs) and light-load hours (LLHs). Heavy-load hours are hours from 06:00 until 22:00 Pacific Time, excluding Sundays and six

North American Electric Reliability Council holidays: New Year's Day, Christmas Day, Labor Day, Thanksgiving Day, Memorial Day, and Independence Day [11]. All other hours are LLHs. The energy rates mentioned in [12] apply to electricity supply at many BPA customer utilities in the Northwest. Also for most utilities that are BPA customers, the demand rates also apply. The demand billing determinant is calculated as the highest hourly power purchased amount during an HLH in a calendar month, less the average power purchased during all HLHs in the month, less a grandfathered amount that has been determined from the particular utility's performance in prior years. Demand charges apply in any month in which the calculated billing determinant is a positive value. The first two terms may be known from a utility's hourly distribution power data for the month. The first two terms are most important for assessing a differential impact from demand charges. The third term is important for calculating a month's demand charges, but it may often be ignored when comparing alternative scenarios. More detailed analysis is necessary if the determinant is often near to and less than zero, which today is not the case for the City of Ellensburg.

## IV. WIND TURBINE PERFORMANCE ANALYSIS

The City of Ellensburg hoped to supplement its power and energy requirements (effectively reducing its demand) with the several wind generators located at its Renewable Energy Park. Following is the analysis of these renewables based on the detailed data collection that consisted of metrology data and the amount of energy produced in five-minute intervals.

### A. Honeywell WindTronics 1.5 kW Model WT6500

This is among the set of five residential-class wind systems tested by the city. This turbine has a unique design with the generator's stator and rotor located distal from the turbine's hub. Electrical generation from this wind turbine stopped March 15, 2013, and was not restored. A wing failed due to an object (perhaps a bird) passing through the spoked generator wheel, which bent it enough to prevent it from rotating. The unit was de-energized as repair parts were not available. In a November 1, 2013, city report, the cause was attributed to "wing failures" [9]. The cost of the system and its components was $16.0 K per year on an annualized basis. Energy data was available from mid-November 2012 until mid-March 2013 when the system failed. The power generation data from this period is discretized because the raw data was reported to the nearest watt-hour each 5-minute interval. Maximum generation seldom, if ever, approached the nameplate value of the wind turbine—1.5 kW. Data was received during only five calendar months. The monthly energy generation and the values of these quantities of energy, according to BPA load-shaping rates, are miniscule. Even if the generator had operated similarly for a full year, all the energy produced over a year would be expected to displace no more than about $1 of energy that the City of Ellensburg would otherwise have purchased. The project analyzed the impact that this asset system had on peak demand; the value was inconsequential. The typical power generation during peak hours was small. Because the resource is intermittent, generation was unlikely to be coincident with a month's peak demand hour.

### B. Home Energy International 2.25 kW Energy Ball® V200

The City of Ellensburg further complemented its power and energy requirements with the power generated by a 2.25 kW Energy Ball wind generator. This is among the set of five residential-class wind systems tested by the city. The annualized cost of the system and its components is estimated at $15.3K. The City of Ellensburg monitored and submitted data from this wind generation system from the middle of September 2012 through October 2013. It remained functional until the city opted to stop testing wind systems altogether and removed it. The

energy generated in each 5-minute interval was recorded, and the project converted this data to average power for each interval. Although the generator capacity is rated at 2.25 kW, it never appears to have generated more than about 0.9 kW during the project period. The value of the energy has been estimated using the HLH and LLH BPA load-shaping rates. Based on the 14 project months that this wind generator was monitored, it should be expected to generate $153 \pm 10$ kWh and thereby displace only about $\$4.16 \pm 0.35$ worth of electrical energy supply during a year, based on the BPA load-shaping rates and the energy supply that was displaced by the generator each month. The impact of this wind generation system on peak demand was evaluated, but it was determined that the impact was inconsequential—less than $1 per year.

### C. Southwest Windpower 2.4 kW Skystream 3.7®

The City of Ellensburg still further complemented its power and energy requirements with the power generated by a 2.4 kW Southwest Windpower Skystream 3.7 wind generator This is among the set of five residential-class wind systems tested by the city during the project. The Skystream 3.7 wind system remained functional until the city opted to stop testing wind systems altogether and removed it in late 2013.In a report dated November 1, 2013, posted on the city's website, the one-time cost of the Skystream system was $24,770 [9]. The data became available for the period from late August 2012 into October 2013. The power achieved and even exceeded the 2.3 kW nameplate rating many times over this period.

All the project's 5-minute power data for this wind system was plotted against the corresponding wind speed data from the 85-foot metrology tower at the renewable energy park near the turbines as in Figure 2. The error bars represent the range of power measurements from the 16th to 84th percentile at each wind speed. The wind speeds were found to have been discretized at the plotted wind-speed magnitudes.

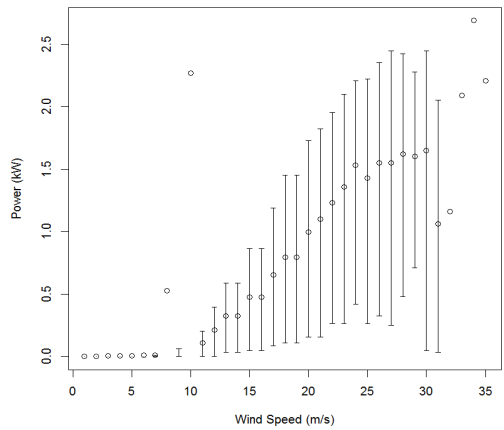

Figure 2. Power generated by the Skystream 3.7 system as a function of wind speed at 85 feet

The characteristic curve is informative, but the project does not believe the wind speed sensors to have been thoroughly calibrated to ensure wind speed accuracy. Nighttime and morning wind generation are less reliable than afternoon.

Figure 3 shows the impact of these diurnal generation patterns on the total monthly HLH and LLH energy. For those several calendar months, for which the project collected data for more than one year, the standard error bars have been included. It would be interesting to learn whether a longer data collection period would smooth the pattern that is observed here during spring and summer months.

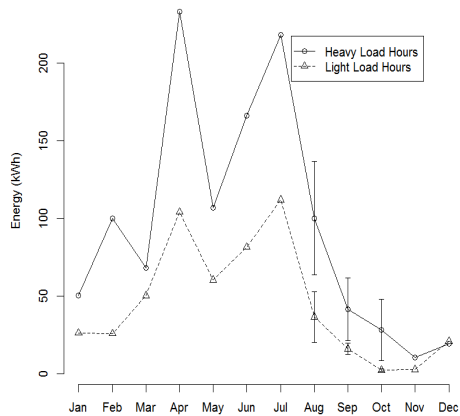

Figure 3. HLH and LLH energy generated each month by the Skystream 3.7 generator

Energy generation each calendar month was compiled and is reported in Table I. Based on the monthly energy summary in this table, the wind generator system should be expected to generate about $1.7 \pm 0.2$ MWh per year. The annual value of this energy, based on the unit costs of HLH and LLH energy that would otherwise be supplied by BPA at its load-shaping rates is $45.13 \pm 6.44$. The city's analysis led them to conclude that this turbine system would produce 1.33 MWh per year, somewhat less than what the project concluded [9]. As has been shown for other wind systems being tested by the City of Ellensburg, the impact of the generation on peak demand and demand charges is negligible. The monthly impacts are summarized in Table I. Negative dollar values in this table represent reductions in the estimated demand charges that the city would incur. The sum impact is an increase in the total yearly demand charges by $0.60 \pm 2.00$.

Table I.
Typical energy generated each month and the monetary value of the displaced energy for the Skystream 3.7 system

| Month | Energy (kWh) | Energy Value ($) |
|---|---|---|
| January | 77 | 2.71 |
| February | 126 | 4.48 |
| March | 118 | 3.32 |
| April | 338 | 8.11 |
| May | 167 | 3.03 |
| June | 248 | 4.97 |
| July | 330 | 9.4 |
| August | $137 \pm 80$ | $4.40 \pm 2.63$ |
| September | $58 \pm 41$ | $1.84 \pm 1.37$ |
| October | $31 \pm 40$ | $0.96 \pm 1.26$ |
| November | 13 | 0.5 |
| December | 40 | 1.5 |

D. *Bergey WindPower 10 kW Excel 10*

The City of Ellensburg further complemented its power and energy requirements with the power generated by a 10 kW Bergey WindPower Excel 10 wind generator located at its Renewable Energy Park. This is one of the four commercial-class wind systems tested by the city. The system was functioning well at the time the city chose to remove all the site's wind generator systems in late 2013. In a report dated November 1, 2013, posted on the city website, the one-time cost of the Bergey system was stated as $96,350 [9]. Data was collected for almost a year, from mid-November 2012 until the end of October 2013. The system achieved and even exceeded its 10 kW nameplate power generation capacity often during this data collection period. The project calculated the characteristic power generated by the Bergey wind generator as a function of wind speed. Figure 4 shows the result of this calculation. All the project's 5-minute interval power data was used along with the corresponding wind speeds that were measured 85 feet above ground at the project's metrology tower at the renewable energy park near the turbine. The markers are at the average generated power for the given wind speeds. The error bars represent the range of power data from the 16th

through the 84th percentile. Wind speeds were found to have been discretized at 1 m/s increments.

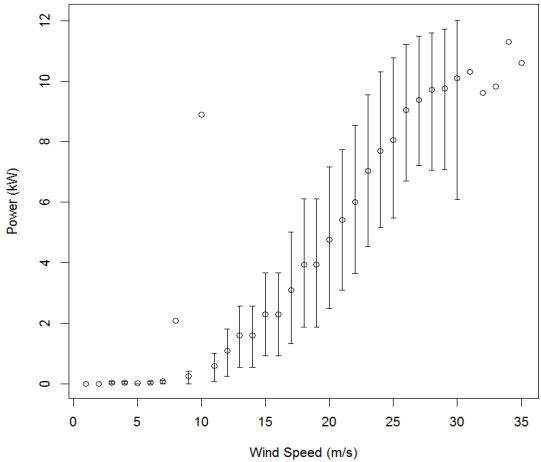

Figure 4. Characteristic power generation of the Bergey system as a function of wind speed

Power generation was much more predictable in the months other than winter. Peak generation often occurred during afternoon hours. Using the average power generation during HLHs and LLHs each month and the numbers of these hours each month of 2013, the project calculated the HLH, LLH, and combined energy generation for each calendar month. Based on the almost full year of operation that is represented in Table III, this generator would generate 7.1 MW and thereby displace $191 worth of supply energy that the city would have otherwise purchased from BPA. The variability of this estimate from one year to another cannot be estimated well because less than one year of data was collected from the operation of this generator. In summer months, the operation of the wind turbine tended to reduce the determinant on which demand charges are applied, but the overall yearly net impact of the generation on incurred demand charges was estimated to be only $0.13.

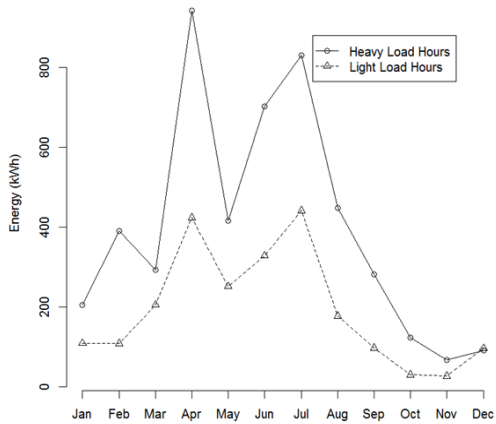

Figure 5. HLH and LLH energy generated each calendar month by the Bergey Windpower system

**Table III.**
**HLH and LLH energy generation for each month and the value of the energy supply that it displaced for the Bergey Windpower system**

| Month | Energy (kWh) | Energy Value ($) |
|---|---|---|
| January | 315 | 11.11 |
| February | 498 | 17.7 |
| March | 497 | 13.96 |
| April | 1360 | 32.8 |
| May | 666 | 12 |
| June | 1030 | 20.7 |
| July | 1270 | 36.07 |
| August | 623 | 19.96 |
| September | 378 | 12.15 |
| October | 153 | 4.72 |
| November | 94 | 3.2 |
| December | 188 | 6.8 |

The use of wind generation neither helps nor hurts the calculation of demand charges throughout the entire year, as summarized in Table IV for each calendar month.

**Table IV.**
**Estimated monthly impact on peak demand for the Bergey Windpower system**

| Month | Δ Demand (kW) | Δ aHLH (kWh/h) | Δ Demand Expense ($) |
|---|---|---|---|
| January | 0.5 ± 2 | 0.5 | 0.19 |
| February | 0.8 ± 2.1 | 1 | 2.54 |
| March | 0.6 ± 1.9 | 0.7 | 1.31 |
| April | 1.0 ± 2.3 | 2.3 | 9.82 |
| May | 1.4 ± 3.1 | 1 | −2.36 |
| June | 2.6 ± 4.9 | 1.8 | −5.55 |
| July | 2.1 ± 4.0 | 2 | −1.10 |
| August | 1.5 ± 3.5 | 1 | −4.52 |
| September | 1.1 ± 2.3 | 0.7 | −4.15 |
| October | 0.1 ± 0.6 | 0.3 | 1.34 |
| November | 0.0 ± 0.3 | 0.2 | 1.30 |
| December | 0.1 ± 0.6 | 0.2 | 1.31 |

*E. Urban Green Energy 4 kW Wind Turbine*

The City of Ellensburg further complemented its power and energy requirements with the power generated by a 4 kW Urban Green Energy wind generator located at its renewable park. This is one of the four commercial-class wind systems tested by the city. The Urban Green Energy generator failed. The unit had a bearing failure. The total system cost $26.8K per year on an annualized basis. Usable data was gathered and delivered for a period from mid-July 2012 until mid-March 2013, a period of about eight months. The generated power magnitudes in 2012 were greater than those in 2013. While this variability could potentially be caused by natural wind variability, it is more likely that the performance of the wind generator degraded over the months that it was monitored by the project. The reduction in wind power was not similarly observed for other of the Ellensburg wind turbines. The initial 2012 power had at times approached the turbine's nameplate capacity, but it never did so in 2013. The impacts of the diurnal generation patterns affected the average monthly HLH and LLH energy generation. Data was available for only eight of the 12 calendar months. Annual generation would be 1.2 MWh, presuming the generator had operated throughout the year and the months of data collection

were typical. The total annual value of displaced energy supply would be $26.93 if the generator had produced throughout the year and presuming the months of data collection can be used to represent the entire year. The total impact of this renewable generation on peak demand might be a reduction of only about $2, based on the impacts during the nine calendar months that data was available to the project.

*F. Ventera Wind 10 kW VT10 Wind Turbine*

The City of Ellensburg installed and operated a 10 kW Ventera Wind VT10 wind generator at its renewable park. This is one of four commercial-class wind systems that were tested by the city. Again, the city was investigating how its need for energy supply and demand might be impacted by renewable energy generation. In a report to its city council, the one-time installed cost of the Ventera Wind system was stated as $110,660 [9]. The City of Ellensburg supplied a data stream of the energy generated every 5 minutes for the period from the beginning of January until late October 2013. The system never generated its 10 kW nameplate capacity, but it occasionally exceeded an 8 kW power output. According to the Ventera Wind website, the generator is rated to start at winds of 2.7 m/s (6 mph) and generate 10 kW at wind speeds of 13 m/s (29 mph). It is advertised to withstand winds at 58 m/s (130 mph). As expected, substantial power was generated at high winds. If the 10 monitored months of generation can be meaningfully extrapolated, then the wind system might generate about 7.2 MWh per year. For comparison, the Ventera Wind website projected that this system would generate 24 MWh per year given a 6.5 m/s (14.5 mph) wind speed at its hub. Extrapolating from the 10 calendar months for which data was available, the value of annual displaced energy supply would be about $192, based on recent BPA load-shaping rates. The variability in this projection will not be estimated because less than one year of data was available. If we can extrapolate the estimated monthly change in demand charges, the result would be a reduction in these charges of only $2.65 for the year. Wind generation does not significantly affect calculated BPA demand charges at this location.

Three more distributed wind system namely, Wing Power 1.4 kW Wind Turbine, Tangarie Gale®(c) and Windspire® 1.2 kW are not discussed here in detail due to space constraints. Windspire® 1.2 kW suffered from generator inverter failure and total value of the energy in monitored months was less than $1.00. The total projected value of the yearly generated energy would be on the order of $2.50 for Tangarie Gale®. The wind turbine's tower was blown over. The city apparently needed to replace the Wing Power wind system several times during the project. In a November 1, 2013, report to the city council, two wing failures had occurred, and failed bolts were recurring issues. The total value of annual generated energy by Wing power is projected to be only about $7.50 ± 2.40. The impact of the Wing power on peak demand was found to be trivial.

## V. PV System Performance Analysis

*A. Thin-Film Solar Panel 54 kW Array*

During the project, the City of Ellensburg added 40.5 kW of nameplate generation capacity to its existing 13.5 kW thin-film PV power generation, bringing the thin-film PV capacity to 54 kW. As with the other renewable generation at this site, the city installed this resource to reduce demand from its energy supplier. The City of Ellensburg submitted the energy that was generated every 5 minutes for a period from the beginning of July 2012 to the end of the project's data collection at the end of August 2014. The project converted these data into average power for the 5-minute intervals. The City of Ellensburg evaluated their Phase-4 installation of additional thin-film PV generation and concluded that, from the city's perspective, the unit cost of the energy produced by the solar system was $0.28/kWh. This is expensive compared to relatively inexpensive wholesale electricity in the Pacific Northwest. The remainder of this section presents independent analysis conducted by the project concerning the entire array of thin-film PV at the Ellensburg renewable energy park.

Table V estimates the monthly energy generated by this system and the value of the energy according to current BPA load-shaping rates. The calculations of yearly energy production were based on average power generation during each calendar month's HLHs and LLHs. That is, the energy reported by the project is not simply the sum energy that was reported to the project for a given time period. The calculated energy is a statistical result that is based on average power generation each calendar month, which may include data from multiple years. This approach may overstate generation somewhat when data were unavailable. Often the loss of data may be attributable to power metering or data collection, but the solar generator operated well. Based on the data in Table V, total annual energy generation is estimated by the project to be 80.7 ± 3.0 MWh.

The total annual HLH and LLH energy usages are expected to be 68 ± 3 MWh and 12.9 ± 0.6 MWh, respectively. The annual displaced energy supply costs for HLHs and LLHs are $2,031 ± 92 and $305 ± 18, respectively, with a total displaced supply value of about $2,335 ± 94 per year.

**Table V.**
**Displaced HLH and LLH supply energy consumptions and costs, based on BPA load-shaping rates for the thin-film PV array**

| Month | Energy (kWh) | Energy Value ($) |
|---|---|---|
| January | 2,600 ± 440 | 97 ± 16 |
| February | 4,700 ± 810 | 171 ± 30 |
| March | 6,800 ± 230 | 201 ± 6 |
| April | 8,900 ± 1,500 | 222 ± 39 |
| May | 9,100 ± 520 | 181 ± 11 |
| June | 9,300 ± 520 | 197 ± 12 |
| July | 9,700 ± 530 | 286 ± 16 |
| August | 8,900 ± 220 | 295 ± 7 |
| September | 7,700 ± 1,100 | 250 ± 37 |
| October | 6,700 ± 1,300 | 208 ± 41 |
| November | 3,500 ± 1,100 | 123 ± 39 |
| December | 2,800 ± 830 | 106 ± 32 |
| Year Total | 80,700 ± 9,100 | $2,337 ± 286 |

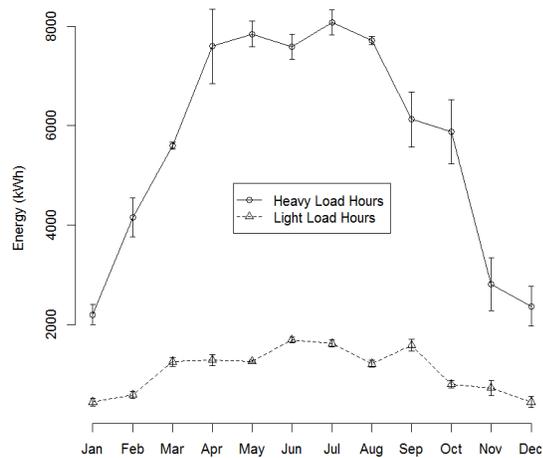

Figure 6. Average HLH and LLH energy generated each month by the PV generator

Table VI.
Estimated monthly impact on peak demand for the thin-film PV array

| Month | Δ Demand (kW) | Δ aHLH (kWh/h) | Δ Expense ($) |
|---|---|---|---|
| January | 0.6 ± 1.7 | 5.3 ± 1.0 | 52 ± 2 |
| February | 4.2 ± 7.7 | 11 ± 2 | 71 ± 8 |
| March | 4.6 ± 6.8 | 13.5 ± 0.3 | 80 ± 7 |
| April | 18 ± 1 | 18 ± 4 | 2 ± 12 |
| May | 24 ± 13 | 19 ± 1 | −31 ± 13 |
| June | 23 ± 14 | 19 ± 1 | −26 ± 14 |
| July | 39 ± 15 | 19 ± 1 | −173 ± 15 |
| August | 26 ± 13 | 17.9 ± 0.4 | −83 ± 13 |
| September | 25 ± 14 | 16 ± 3 | −94 ± 14 |
| October | 4.2 ± 6.3 | 14 ± 3 | 87 ± 7 |
| November | 0.4 ± 1.9 | 7.0 ± 2.7 | 70 ± 3 |
| December | 0.0± 0.4 | 5.9 ± 2.0 | 67 ± 2 |

The generated HLH and LLH energies for each calendar month have been plotted in Figure 6. This figure emphasizes that most of the energy is generated during HLHs. Nearly four times more energy is generated in the summer when the sun is high in the sky than is generated during winter months, when the sun remains low on the horizon and is often hidden by clouds. The summer peak power generation is approximately twice that of winter. The productive summer day includes more morning and evening hours than in other seasons. Power generation in the summer is more predictable than for other months. The error bars are demand-charge determinant from which a change in monthly demand charges may be calculated. The city incurs demand shorter. The variability of generation is somewhat greater in the afternoon than during the morning hours. Table VI presents the change in aHLH and the change in peak-hour demand that may be attributed to this thin-film system's power generation. It was used to estimate typical hours of the city's peak electrical load. The normal generation and variation of generation for these hours were used to estimate the impacts during peak power during HLHs each month with an estimate of how this average might change from year to year. Finally, the total hours. The aHLH component is simply the average generated monthly impact was estimated from these analysis results and These are the two main components of the charges nearly every month. The project received from Ellensburg a history of their peak hours. The most recent 24 of these examples (i.e., two years' worth) were the BPA demand rate . This monthly impact is shown in Table VI. The negative values that occur primarily during summer months represent reductions of the city's demand charges those months. The demand charges are increased for many other months because the solar power generation is unimpressive during the (typically) morning peak hours those months. The total impact of the solar generation from the thin-film PV system on the municipality's demand charges is only $22 ± 36 per year. Note that this is an overall increase in the demand charges the city would pay. Regardless, the magnitude is relatively insignificant.

## VI. CONCLUSIONS AND DISCUSSION

This paper discusses the renewable energy sources at the Ellensburg Community Renewable Park, Washington. Residents of Ellensburg could purchase shares in the energy production of the generators at this community park. The municipality installed, maintained, and completed distribution connectivity of these generators for the residents. It thereby consolidated renewable resources that might otherwise be integrated on different location of distribution feeders, house or rooftops. Following detailed technical and policy observations were made based on this analysis.

The experiment with distributed wind turbines encountered a number of challenges. City of Ellensburg installed additional capacity of 40 kW to its existing PV installation. The results for this particular installation which happened in conjunction with distributed wind are presented here. The project analyzed data from the summer of 2012 until spring of 2014. The overall PV capacity factor was 34.5% and energy production was 173 MWh. The lowest energy production was in winter of 2013 (18.9 MWh) and the capacity factor for the winter of 2013 was 24.6%. Whereas, highest energy production was in summer of 2013 (27.5 MWh) and the corresponding highest capacity factor was 38.7%. The City of Ellensburg Renewable Energy Park installed 9 small and medium scale distributed wind turbines. Total installed wind capacity was 43.6 kW. The following table summarizes capacity factors and energy production from these 9 wind generators. Five of the nine wind turbine systems had failed by the time the city removed them in fall 2013. After a turbine tower collapsed, the city resolved that wind turbines should not operate so close to residential foot traffic in the Renewable Energy Park. The turbine systems' seasonal capacity factors were quite low. The systems having greater nameplate capacities typically achieved significantly better capacity factors than did the smaller, residential-scale turbine systems.

**Table VII.**
**Projected annual energy production and capacity factors of tested small wind turbine models**

| Wind Turbine Make and Model | Energy Production (kWh) | Capacity Factor (%) |
|---|---|---|
| Honeywell WindTronics WT6500[a] | 29 | 0.2 |
| Windspire® v1.2[a] | 76 | 0.7 |
| Home Energy International Energy Ball® V200 | 153±10 | 0.8 |
| Southwest Windpower Skystream® 3.7 | 1700±200 | 8.1 |
| Wing Power Energy | 270±80 | 2.2 |
| Bergey WindPower Excel 10 | 7100 | 8.1 |
| Tangarie Gale®[a] (insufficient data) | - | - |
| Urban Green Energy[a] | 1200 | 3.4 |
| Ventera VT10 | 7200 | 8.2 |

[a] Failed during demonstration period

The levalized cost of energy (LCOE) is measure of the per-kilowatt hour cost of operating a power plant over its lifetime. A small LCOE is essential from the customer point of view. In this study, only two out of nine distributed wind projects performed relatively better. Most of the residential class distributed wind generators failed to give long term robust performance. Maintenance issues in some cases were apparent. This increased the operation and maintenance cost. Robust, low-maintenance designs of distributed wind turbines can help to drive LCOE lower.

For one of the wind turbines, a tower structure failed under high winds. Several others suffered from various mechanical break downs. There were also instance of vandals destroying some electrical components. Physical security of renewable energy parks is a challenge that needs to be addressed for successful adoption of the renewable energy park concept. Zoning authorities across the U.S. are sometimes concerned about the safety impacts of distributed turbines. As per PNWSGD experience, their concerns are valid.

Careful examination of the small distributed wind suppliers is necessary before purchase. All distributed wind turbines installed in this project were in years 2012-13. The customers may leverage American Wind Energy Association (AWEA) evaluation process. Effective from February 2015 through December 2015 in order to qualify for federal tax credits, AWEA certification is mandatory. It is worth noting that Bergey WindPower 10 kW Excel 10 and Southwest Windpower 2.4 kW Skystream which are certified by AWEA (as on November 2015) remained functional and performed relatively well throughout the PNWSGD project timeline. For, others robust long term performance of distributed wind came across as a critical bottleneck.

The objective of the project to reduce our future BPA Tier 2 obligations was not achieved due to failure of distributed wind technology. In most of the cases, annual energy production and economic benefits were low. Months after installation, several turbines entirely stopped producing energy. To ensure quality of integrated renewables, performance based incentive type structure will be useful rather than federal and state incentives based on a system's rated capacity. The performance goals which were promised by manufactures were not achievable with many of the tested wind systems. The PV arrays, on the other hand, performed relatively well.

The City of Ellensburg Renewable Energy Research Park has demonstrated the viability of the community energy park concept. Before wide-scale adoption of any renewable energy technology, renewable energy parks offer a means for citizens to invest in and test available renewable generation options. Other allied objectives such as education, eco-tourism, and research and technology validation can be achieved. This paper has summarized our experience with both performance and cautionary issues for others who might be considering this interesting option.

## VIII. BIOGRAPHIES

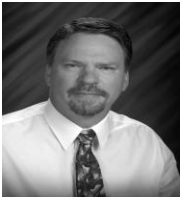

**Donald J. Hammerstrom** (S'94, M'95, SM'06) earned a B.A. in chemistry from St. Olaf College, Northfield, Minnesota, a B.S. in education from Eastern Montana College, Billings, Montana and M.S. and Ph.D. degrees in electrical engineering from Montana State University, Bozeman, Montana in 1991 and 1994, respectively. He is employed by Battelle Memorial Institute as a senior research engineer in the Energy and Energy Division of Pacific Northwest National Laboratory, Richland, Washington. He has managed field demonstrations of smart grid technologies for the U.S. Department of Energy, leads technology development of grid-responsive loads, and develops power electronic converter applications. Prior to joining Battelle, he designed power converters, biological sample collectors, and surface decontamination systems for startup companies in Washington State. He has authored United States patents in the diverse areas of energy management systems, power electronic converters, microtechnology, microbe decontamination, and aerosol sample collection.

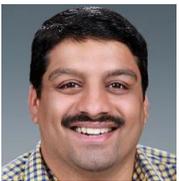

**Yashodhan P. Agalgaonkar** (M'14) received the M.Sc. degree in Electrical Power Engineering from the Chalmers University of Technology, Gothenburg, Sweden, in 2006, and the Ph.D. degree in Electrical Power Engineering from the Imperial College London, London, U.K., in 2014. He was a Postdoctoral Researcher at the Imperial College London until 2014. From 2006 to 2010, he was with Crompton Greaves, India and with Converteam (now GE Energy), Chennai, India and Berlin, Germany as a Research Engineer. During the tenure at Crompton Greaves and Converteam, he conducted research on diverse areas in Power Transmission and Distribution operation. Presently, since 2015 he is Scientist and Engineer in the Energy and Energy Division of Pacific Northwest National Laboratory, Richland, Washington.

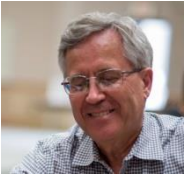

**Stephen E. Elbert** (M'00) joined Pacific Northwest National Laboratory (PNNL) in 2004 as a computational scientist. He has been a program manager at the Department of Energy and the National Science Foundation. His research interests are scalability, efficiency, and productivity of high-performance computing systems, especially as applied to electric power grid related problems such as AC optimal power flow for security constrained unit commitment, financial transmission rights, and dynamic state estimation using ensemble Kalman filters. He led the data collection effort for PNWSGD. He received a Ph.D. in computational chemistry from the University of Washington. He is a member of the IEEE, the American Association for the Advancement of Science, and Sigma Xi.